\documentclass[prd,aps,10pt,nofootinbib,twocolumn,superscriptaddress,preprintnumbers,balancelastpage,longbibliography]{revtex4-1}

\usepackage{booktabs}    
\usepackage{placeins}
\usepackage{soul}
\usepackage[usenames,dvipsnames,svgnames]{xcolor}
\definecolor{redd}{rgb}{0.8, 0.1,0.2}
\definecolor{navy}{rgb}{0.05, 0.23,0.75}
\usepackage[colorlinks=true,citecolor=red,linkcolor=blue]{hyperref}
\usepackage{chngcntr}
\hypersetup{
     colorlinks   = true,
     citecolor    = navy,
	linkcolor = redd,
	urlcolor=navy,
	anchorcolor=blue
}
\usepackage{bbm}
\usepackage{amsfonts}
\usepackage{amsmath,amssymb}
\usepackage{mathrsfs}
\usepackage{epsfig}
\usepackage{graphicx}
\usepackage{wrapfig}                 
\usepackage{url}
\usepackage{hyperref}
\usepackage{float}
\usepackage{color}
\usepackage{multirow}
\usepackage{lipsum}
\usepackage{enumitem}
\usepackage{ctable}
\newcolumntype{L}{>{\centering\arraybackslash}m{1.5cm}}

\usepackage{enumitem}

\usepackage{chngcntr}

\newcommand{\bear}{\begin{array}}
\newcommand {\eear}{\end{array}}

\newcommand{\beq}{\begin {equation}}
\newcommand{\eeq}{\end   {equation}}
\newcommand{\bea}{\begin {eqnarray}}
\newcommand{\eea}{\end   {eqnarray}}
\newcommand{\baa}{\begin {array}   }
\newcommand{\eaa}{\end   {array}   }
\newcommand{\bit}{\begin {itemize} }
\newcommand{\eit}{\end   {itemize} }
\newcommand{\be }{\begin {equation}}
\newcommand{\ee }{\end   {equation}}
\newcommand{\nn }{\nonumber        }

\def\bea{\begin{eqnarray}}
\def\eea{\end{eqnarray}}

\newcommand{\bef}{\begin{figure}}
\newcommand {\eef}{\end{figure}}
\newcommand{\bec}{\begin{center}}
\newcommand {\eec}{\end{center}}

\definecolor{cerulean}{rgb}{0., 0.62,0.7}

\newcommand{\twiddle}{{\raise.17ex\hbox{$\scriptstyle\sim$}}}

\begin{document}

\title{Axion Free-kick Misalignment Mechanism}

\author{Ling-Xiao Xu}
\email{lingxiao.xu@unipd.it}
\affiliation{Dipartimento di Fisica e Astronomia `G. Galilei', Universit\`{a} di Padova, Italy}
\affiliation{Istituto Nazionale di Fisica Nucleare (INFN), Sezione di Padova, Italy}

\author{Seokhoon Yun}
\email{seokhoon.yun@pd.infn.it}
\affiliation{Dipartimento di Fisica e Astronomia `G. Galilei', Universit\`{a} di Padova, Italy}
\affiliation{Istituto Nazionale di Fisica Nucleare (INFN), Sezione di Padova, Italy}

\begin{abstract}
We propose an alternative scenario for the axion misalignment mechanism based on the nontrivial interplay between the axion and a light dilaton in the early universe. Dark matter abundance is still sourced by the initial misalignment of the axion field, whose motion along the potential kicks the dilaton field away from its minimum, and dilaton starts to oscillate later with a delayed onset time for oscillation and a relatively large misalignment value due to the kick; eventually the dilaton dominates over the axion in their energy densities, and the dilaton is identified as dark matter. The kick effect due to axion motion is the most significant if the initial field value of dilaton is near its minimum; therefore, we call this scenario axion ``free-kick" misalignment mechanism, where axion plays the role similar to a football player. Dark matter abundance can be obtained with a lower axion decay constant compared to the conventional misalignment mechanism.
\end{abstract}

\maketitle

\noindent
\noindent\textbf{Introduction.---}
Ultralight scalars are ubiquitous in particle physics; they can play various important roles in solving the strong CP problem~\cite{Peccei:1977hh,Peccei:1977ur,Weinberg:1977ma,Wilczek:1977pj}, attempting to solve the cosmological constant problem~\cite{Abbott:1984qf}, generating dark matter (DM) relic abundance~\cite{Preskill:1982cy,Abbott:1982af,Dine:1982ah}, explaining matter-antimatter asymmetry~\cite{Affleck:1984fy}, driving inflation~\cite{Guth:1980zm,Linde:1981mu}, and selecting the weak scale in the early universe~\cite{Graham:2015cka}. (See e.g.~\cite{TitoDAgnolo:2021nhd,TitoDAgnolo:2021pjo} where the hierarchy problem of the weak scale and the strong CP problem are solved jointly.) Ultralight scalars can also arise from string theory~\cite{Svrcek:2006yi} and lead to various interesting phenomenological consequences~\cite{Arvanitaki:2009fg}. In recent years, ultralight scalars are under extensive phenomenological and experimental scrutinies (see~\cite{Antypas:2022asj,Adams:2022pbo} for reviews). Given the richness of possible ultralight scalars, there might be non-trivial interplay between them that can change the conventional picture. 

In the conventional misalignment mechanism~\cite{Preskill:1982cy,Abbott:1982af,Dine:1982ah}, the DM abundance is produced in the early universe due to the initial misalignment of scalar field value away from its minimum. When the time-dependent Hubble parameter drops below the scalar mass, the scalar field starts to oscillate, and its energy density redshifts as non-relativistic matter as the universe expands. The dynamics of scalar misalignment crucially depends on the initial condition, the shape of the scalar potential, and interactions between the scalar and other particles. Along these directions, one can possibly modify the conventional misalignment mechanism, and there has been significant progresses in recent years, especially for axions or more generally axion-like particles (ALPs). Several novel and interesting scenarios are discussed in Refs.~\cite{Co:2019jts,Co:2019wyp,Chang:2019tvx,Huang:2020etx,DiLuzio:2021gos,Choi:2022nlt,Papageorgiou:2022prc,Allali:2022sfm,Allali:2022yvx,Batell:2021ofv,Batell:2022qvr}.
On the other hand, identifying new scenarios of scalar misalignment with interesting phenomenology is still an ongoing endeavor.

\begin{figure}[t] 
\centering 
\includegraphics[width=0.43\textwidth]{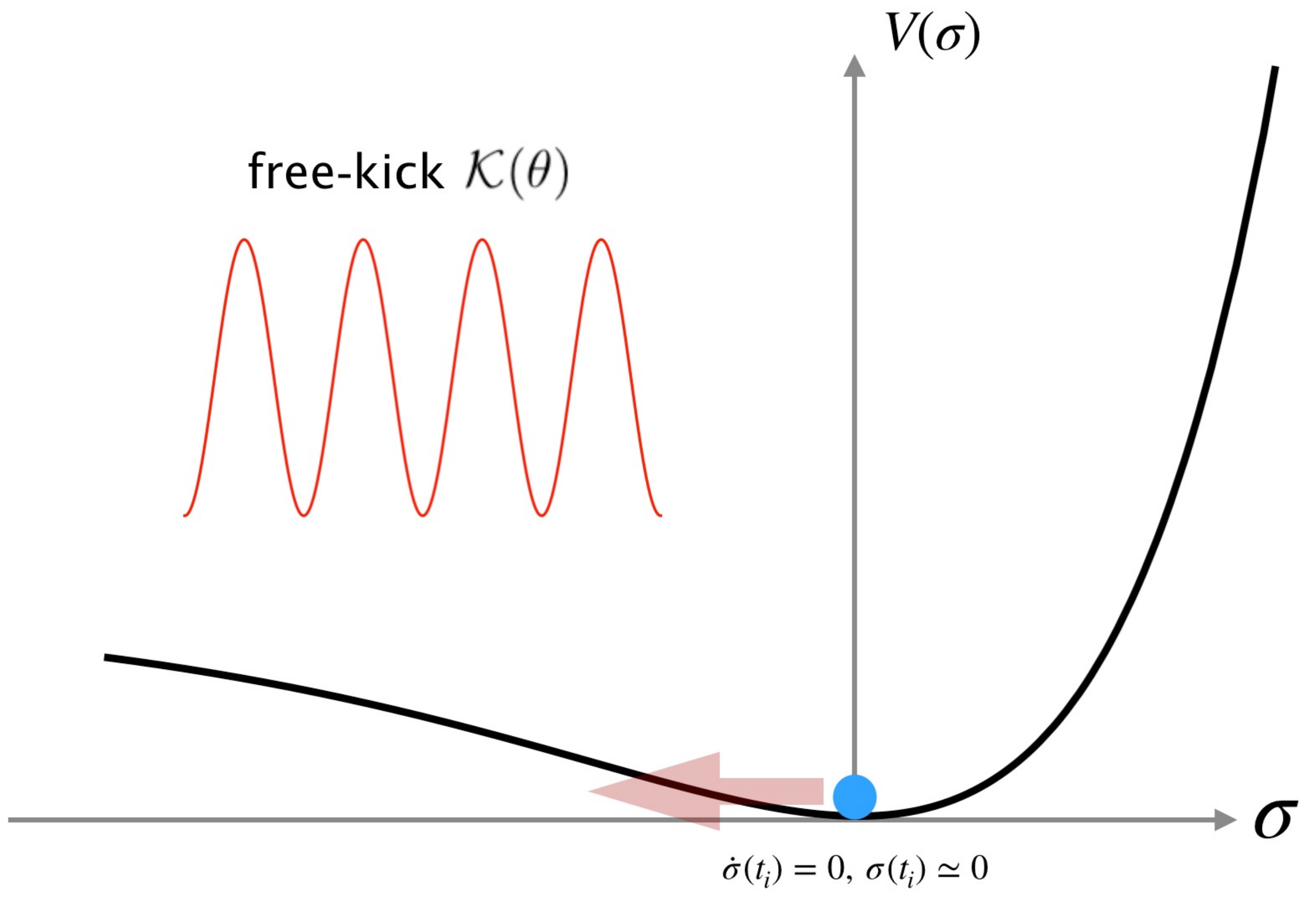} 
\caption{Cartoon of the Axion free-kick misalignment mechanism. We consider the case where axion is heavier than dilaton, such that axion starts to oscillate first. The motion of axion $\theta$ along its potential kicks the dilaton $\sigma$ away from its minimum through the kick term denoted as $\mathcal{K}(\theta)$, where axion plays the role similar to a football player. Later on, the dilaton starts to oscillate with a relatively large field value due to the axion kick and a delayed onset time for its oscillation. Eventually, dilaton energy density dominates and redshifts as cold dark matter.}
\label{fig:sketch} 
\end{figure}

In this Letter, we propose a new alternative to the conventional axion misalignment mechanism~\cite{Preskill:1982cy,Abbott:1982af,Dine:1982ah} and several others~\cite{Co:2019jts,Co:2019wyp,Chang:2019tvx,Huang:2020etx,DiLuzio:2021gos,Choi:2022nlt,Papageorgiou:2022prc,Allali:2022sfm,Allali:2022yvx,Batell:2021ofv,Batell:2022qvr}. 
Our scenario is based on the interactions between an axion and a light dilaton if the sector relevant for the axion is UV completed into a scale invariant theory, whereas at low energy scale invariance (or conformal symmetry) is nonlinearly realized by the dilaton field~\cite{Salam:1969bwb,Isham:1970gz,Isham:1971dv}. In this work, we do not distinguish axion from ALPs, and dilaton in general means an ultralight scalar field with dilatonic couplings (i.e., a dilaton-like particle). In particular, we consider that the axion mass is larger than the dilaton mass, and we assume zero initial velocities for both the axion and the dilaton. 
In our scenario, the DM abundance is still initially sourced by the axion misalignment away from the minimum of axion potential. Therefore, only the initial axion energy density is nonvanishing, whose value is the same as in the conventional misalignment mechanism for any fixed values of axion mass, decay constant, and initial axion field value. When the Hubble parameter drops below the axion mass, axion starts to oscillate first, whose motion kicks the dilaton away from its minimum. Here, the axion plays a role similar to a football player. Later on, when the Hubble parameter drops below the dilaton mass, dilaton starts to oscillate with a relatively large (but still order one) field value due to the axion kick and a delayed onset time for oscillation. Eventually, dilaton dominates over axion in their energy densities, and dilaton is identified as cold DM. The key points of our mechanism are depicted in Fig.~\ref{fig:sketch}, where the kick effect is most efficient if initially the dilaton is trapped near its minimum. In this case, we call it the Axion ``free-kick" misalignment mechanism. Otherwise, DM abundance is also partially sourced by the initial dilaton misalignment, the kick effect is less efficient. Notice that the `kick' effect on ultralight scalars can also be induced by their interactions with other particles in the thermal bath~\cite{Croon:2022gwq}. In this work, we propose the general mechanism and find a particular realization using a light dilaton.

As we will justify, due to the interplay between axion and dilaton, the dark matter abundance can be reproduced with a lower axion decay constant $f_\theta$ in our scenario, which is interesting for all the axion detection experiments.

\vspace{0.1in}
\noindent
\noindent\textbf{Setup.---}
We consider a scenario where the axion theory is UV completed into a scale invariant theory, whereas at low energy scale invariance is non-linearly realized. In the broken phase, the theory can be formally promoted to a scale-invariant one using the Weyl compensator~\cite{Vecchi:2010gj,Chacko:2012sy}
\bea
\chi=F\hat{\chi}= F e^{\frac{\sigma(x)}{F}}\ ,
\eea
where $\sigma$ is the dilaton field with $F$ being its decay constant. For example, scale invariance is formally broken if there is a dimensionful scale or mass $f$ in the theory, as $f\to e^\Delta f$ under scale transformation, where $\Delta$ is a proper parameter. However, one can dress the Weyl compensator $\hat{\chi}$ on $f$ such that the combination $f\hat{\chi}$ is invariant under scale transformation. The dilaton realizes the scale invariance non-linearly, since rescaling the dimensionful parameter $f$ is equivalent to shifting the $\sigma$ field value, i.e. under scale transformation $\sigma\to \sigma-F \Delta$ such that $f\hat{\chi}$ is invariant. Notice that the dilaton $\sigma$ and the Weyl compensator $\hat{\chi}$ are completely analogous to the pions $\pi$ and the pion matrix $U=e^{i \frac{\pi(x)}{f_\pi}}$ in the conventional chiral Lagrangian, which are used to realize the chiral symmetry non-linearly.

In this work, we study the interactions between dilaton $\sigma$ and axion $\theta$ from the viewpoint of effective theories. Following the previous considerations, we have 
\bea
\mathcal{L}=\sqrt{|g|} \left\{ \frac{f_\theta^2 \hat{\chi}^2}{2} (\partial_\mu \theta)^2+\frac{1}{2} (\partial_\mu \chi)^2-V(\sigma, \theta) \right\} \ ,
\eea
where the axion decay constant $f_\theta$ breaks scale invariance explicitly, nevertheless the combination $f_\theta \hat{\chi}$ is formally scale invariant. One can recover the conventional axion kinetic term by taking the limit $\hat{\chi}\to 1$.
As in the conventional case, we assume a radiation-dominated Universe with the flat FRW metric $g=\text{diag}\left(1,-R(t)^2,-R(t)^2,-R(t)^2\right)$ (where the scale factor is denoted as $R(t)$), hence $\sqrt{|g|}=R(t)^3$. 

The full potential $V(\sigma, \theta)$ consists of the dilaton potential $V(\sigma)$ and the axion potential $V(\theta)$, i.e.
\bea
V(\sigma, \theta)&=& V(\sigma) + \hat{\chi}^4\ V(\theta)\nn\\
&=&V(\sigma)+ m_\theta^2 f_\theta^2 \hat{\chi}^4 \left(1-\cos[\theta]\right)\ ,
\eea
where $m_\theta$ and $f_\theta$ are the mass and decay constant of axion. Both $m_\theta$ and $f_\theta$ break scale invariance explicitly, but the combination $m_\theta^2 f_\theta^2 \hat{\chi}^4$ is formally scale invariant. Similar to the axion kinetic term, here the usual axion potential is recovered when $\hat{\chi}\to 1$. In this work, we keep the values of $m_\theta$ and $f_\theta$ being general, while they have to satisfy the constraint $m_\theta^2 f_\theta^2\sim (80 \text{MeV})^4$ for QCD axion.
In this work, we keep the origin of dilaton potential agnostic, it in general can take the form of
\bea
V(\sigma)&=&\lambda F^4 \left(-\frac{4}{4-\epsilon} \hat{\chi}^{4-\epsilon}+\hat{\chi}^4\right)+\lambda F^4 \frac{\epsilon}{4-\epsilon} \\
&=&\lambda F^4 \left(-\frac{4}{4-\epsilon} e^{(4-\epsilon)\frac{\sigma}{F}}+e^{4\frac{\sigma}{F}}\right)+\lambda F^4 \frac{\epsilon}{4-\epsilon}\ , \nn
\eea
where the parameter $\lambda$ controls the overall magnitude of the dilaton potential, and $\epsilon> 0$ parametrizes the explicit breaking of scale invariance, i.e. the operator $F^4\hat{\chi}^4$ is formally scale invariant while the operator $F^4\hat{\chi}^{4-\epsilon}$ breaks scale invariance explicitly. The constant term $\lambda F^4 \frac{\epsilon}{4-\epsilon}$ breaks scale invariance as well, it is added such that $V(\sigma)=0$ when $\sigma=0$.
Roughly speaking, the dilaton potential is of order $\lambda F^4$, which might dominate over the axion potential, whose magnitude is of order $m_\theta^2 f_\theta^2$, or vice versa.
The minimum of $V(\sigma)$ corresponds to $\langle \hat{\chi} \rangle=1$ (or equivalently $\langle \sigma \rangle=0$). Accordingly, the global minimum of $V(\sigma, \theta)$ corresponds to $\langle \sigma \rangle=0$ and $\langle \theta \rangle=0$ mod $2 \pi$. 
The masses of dilaton and axion are $m_\sigma^2=4 \epsilon \lambda F^2$ and $m_\theta^2$, respectively. 
It is natural to require that $F\geq f_\theta$, since scale invariance is necessarily broken once Peccei-Quinn (PQ) symmetry is broken. 
Nevertheless, since $\lambda$ and/or $\epsilon$ can be small, the dilaton can be very light; see e.g.~\cite{Appelquist:2010gy,Coradeschi:2013gda,Bellazzini:2013fga} on how to naturally obtain a light dilaton.

\vspace{0.1in}
\noindent
\noindent\textbf{Mechanism.---}
The equations of motion of $\theta$ and $\sigma$ can be worked out straightforwardly. In particular, $\theta$ and $\sigma$ are assumed to be spatially homogeneous and only time-dependent. 
The equations of motion are
\bea
\ddot{\theta}(t)+\left(3 H+\frac{2}{F}\dot{\sigma}(t)\right) \dot{\theta}(t)+ m_\theta^2 \ e^{2\frac{\sigma(t)}{F}}\sin\left[\theta(t)\right]=0\ , \nn\\
\label{eq:eom_axion}
\eea
and
\bea
&\ &\frac{\ddot{\sigma}(t)}{F}+\left(3 H + \frac{\dot{\sigma}(t)}{F} \right) \frac{\dot{\sigma}(t)}{F} + \frac{m_\sigma^2}{\epsilon} e^{2\frac{\sigma(t)}{F}}\left(1-e^{-\epsilon \frac{ \sigma(t)}{F}}\right) \nn\\
&+&4 \frac{m_\theta^2 f_\theta^2}{F^2} \left(1-\cos[\theta(t)]\right)  e^{2 \frac{\sigma(t)}{F}}- \frac{f_\theta^2}{F^2} \dot{\theta}(t)^2 =0\ .
\label{eq:eom_dilaton}
\eea
There is nontrivial interplay between the evolution of axion and dilaton. 
As seen in Eq.~(\ref{eq:eom_axion}), the dilaton velocity and dilaton field value effectively change the Hubble friction and mass for the axion, respectively. If $\dot{\sigma}(t)>0$ while $\sigma(t)<0$, the effective Hubble friction for the axion is increased and the effective mass of the axion is decreased, so the onset of axion oscillation is delayed in this case. As seen in Eq.~(\ref{eq:eom_dilaton}), the motion of axion can also back-react to the evolution of dilaton via the `kick' term
\bea
\mathcal{K}(\theta)= \frac{4 m_\theta^2 f_\theta^2 \left(1-\cos[\theta]\right) e^{2 \frac{\sigma(t)}{F}} - f_\theta^2 \dot{\theta}^2}{F^2}\ ,
\label{eq:kick_term}
\eea
which kicks the dilaton even when the dilaton is sitting at the minimum of $V(\sigma)$. Notice the `kick' term does not vanish as long as the axion field is not sitting in the minimum of $V(\theta)$ and it is efficient without the exponential suppression from the dilaton, for example, when $\sigma\sim 0$.

The dynamics of misalignment depends on the masses and initial conditions. If $m_\sigma>m_\theta$, the dilaton just starts to oscillate first if the initial misalignment is not zero and then settles in the minimum afterwards. In this case, there is almost no interplay between axion and dilaton; the oscillation of the axion would be the same as in the conventional scenario, despite the fact that the dilaton may still dominate the energy density due to its large mass. Therefore, we only consider $m_\theta>m_\sigma$ instead. For simplicity, we also assume that the velocities are zero for both the axion and dilaton at initial time $t_i$, i.e. $\dot{\theta}(t_i)=\dot{\sigma}(t_i)=0$.  
The axion free-kick mechanism is efficient when the initial misalignment of the axion is nonzero while that of dilaton is zero, i.e. 
\bea
\theta(t_i)\sim \mathcal{O}(1),\ \sigma(t_i)\simeq 0\ .
\label{eq:initial}
\eea
Notice that $\sigma(t_i)\simeq 0$ can be dynamically realized if the effective dilaton mass is enhanced during inflation, such that, if the effective dilaton mass is larger than the inflationary Hubble scale, the dilaton field value is relaxed to zero naturally. Similar idea was proposed in Ref.~\cite{Co:2018phi} to dynamically realize a small initial axion misalignment value.
In the case of Eq.~(\ref{eq:initial}), only the initial axion energy density is nonzero, which sources the DM abundance; otherwise, the DM abundance is sourced by both the initial misalignment of the axion and the dilaton, and the kick effect can be less important even though it does not vanish. For example, when $m_\theta>m_\sigma$ while $\sigma(t_i)/F\sim \mathcal{O}(-1)$, the onset of axion misalignment is delayed due to the dilaton-dependent exponential suppression for the effective axion mass, this is different from the conventional axion misalignment scenario. However, the kick effect is also suppressed due to the same exponential factor. We find numerically that the total DM abundance is also enhanced with dilaton dominance, but a detailed analysis is beyond the scope of this work. 

Despite the fact that DM abundance is still initially sourced by axion misalignment, the role of axion in our scenario is different from other previous scenarios, which makes our mechanism distinct. 
When the Hubble parameter $H$ drops below the axion mass $m_\theta$, axion starts to oscillate and kicks the dilaton away from its minimum via $\mathcal{K}(\theta)$ in Eq.~(\ref{eq:kick_term}); here axion plays the role similar to a football player. Through the kick, axion transfers its energy density to dilaton. When the Hubble parameter is still above the dilaton mass $m_\sigma$, the dilaton misalignment gets accumulated until the onset of its oscillation, which happens when the Hubble drops below the dilaton mass. The amount of dilaton misalignment at the onset of its oscillation is determined by the kick effect, which is more significant if the ratio $m_\theta/m_\sigma$ is bigger. Moreover, a smaller $m_\sigma$ also delays the onset of oscillation. After the onset of dilaton oscillation, its energy density dominates over that of the axion and it redshifts like matter. 
Eventually, we will have dilaton DM in our scenario. 

\begin{figure}[t] 
\centering 
\includegraphics[width=0.5\textwidth]{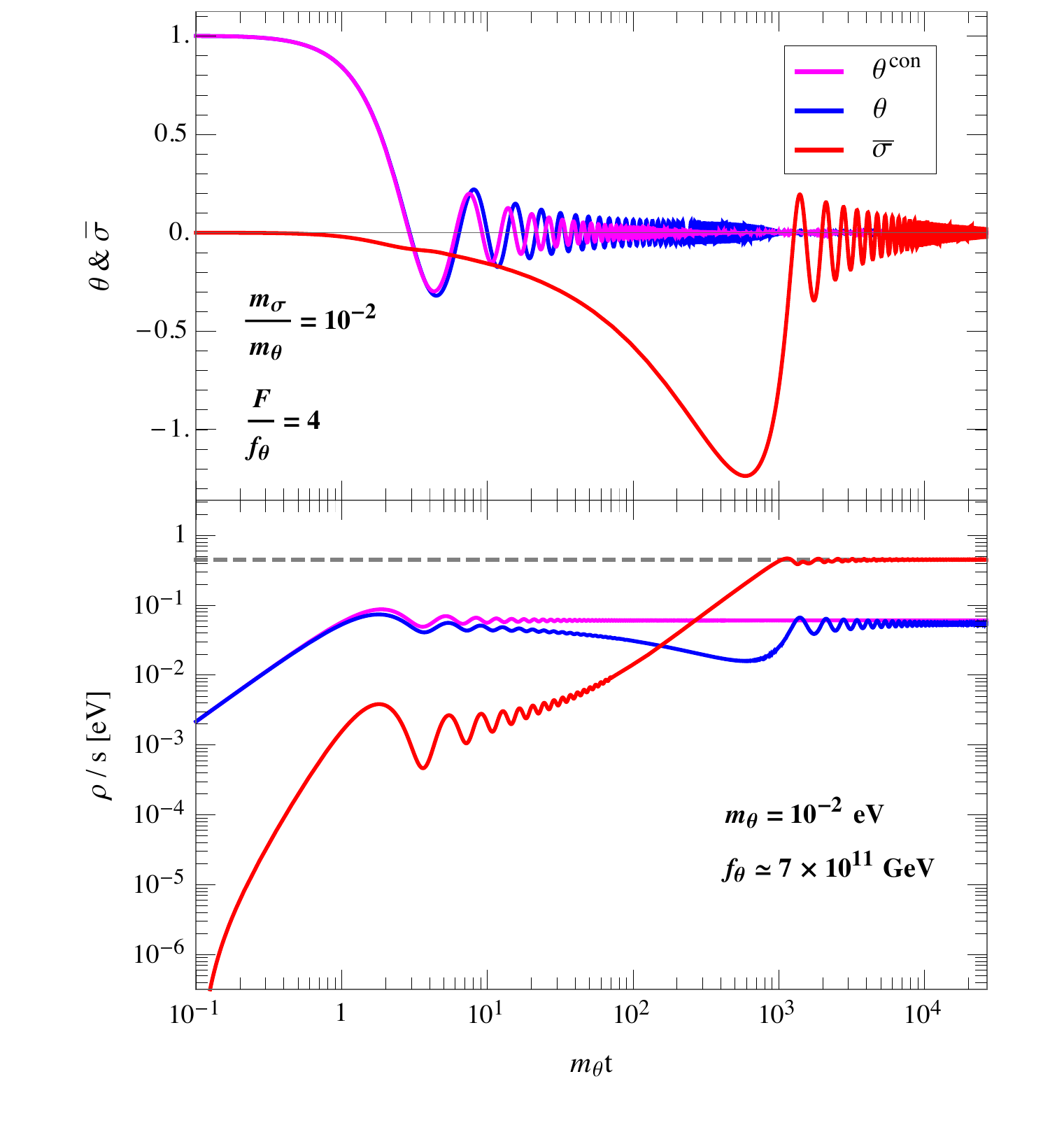} 
\caption{Time evolution of dimensionless axion $\theta$ and dilaton $\bar\sigma\equiv \sigma/F$ field values (upper panel) and their energy densities normalized with the entropy density $s$ (lower panel), where the axion and dilaton in our free-kick misalignment mechanism are shown by the blue and red lines, respectively. As a benchmark, we choose parameters $m_\theta=10^{-2}$ GeV and $f_\theta\simeq 7\times 10^{11}$ GeV for the axion mass and decay constant, while the dilaton mass and decay constant can be read off through the relations $m_\sigma/m_\theta=10^{-2}$ and $F/f_\theta=4$.
The axion in the conventional misalignment mechanism with the same $m_\theta$ and $f_\theta$ is denoted as $\theta^{\text{con}}$, whose evolution is described by the magenta line.  The observed value of comoving DM energy density is shown by the dashed line (lower panel).}
\label{fig:evolution} 
\end{figure}

The time evolution of axion and dilaton in our mechanism is depicted in the upper panel of Fig.~\ref{fig:evolution}, which is shown by the blue and red lines, respectively. 
For comparison, the motion of axion (with the same mass and decay constant) in the conventional misalignment mechanism is also shown by the magenta line. 
For convenience, here and in the following, we use the dimensionless dilaton field, which is defined as $\bar\sigma\equiv \sigma/F$.

\vspace{0.1in}
\noindent
\noindent\textbf{Dark Matter Abundance.---}
The energy densities of the axion and the dilaton are
\bea
\rho_\theta=\frac{f_\theta^2 \hat{\chi}^2}{2} \dot{\theta}^2+\hat{\chi}^4\ V(\theta)\ ,
\label{eq:ed_axion}
\eea
and
\bea
\rho_\sigma=\frac{1}{2} \dot{\chi}^2+V(\sigma)\ ,
\label{eq:ed_dilaton}
\eea
respectively. The total energy density is given by $\rho=\rho_\theta+\rho_\sigma$, it is useful to define the redshift-invariant quantity $\rho/s$, where $s(T)=\frac{2\pi^2}{45} g_{\text{eff}} T^3$ is the entropy density and $R(T) \ T=\text{constant}$. We fix $g_{\text{eff}}\sim 100$ throughout the calculation, but the result does not crucially depend on the chosen value of $g_{\text{eff}}$. 
The comoving energy densities $\rho_\theta/s$ and $\rho_\sigma/s$ are shown in the lower panel of Fig.~\ref{fig:evolution}, where $\rho_\theta/s$ is shown by the blue line, and $\rho_\sigma/s$ is shown by the red line. For comparison, we also include the axion comoving energy density (with the same axion mass and decay constant) in the conventional misalignment mechanism, which is shown by the magenta line. 
The observed comoving DM energy density is indicated by the dashed horizontal line, which is $\rho^{\text{obs}}/s\simeq 0.45$ eV. 
The numerical values of $m_\theta$, $f_\theta$, $m_\sigma$, and $F$ are chosen only as a benchmark. 
One can easily see from Fig.~\ref{fig:evolution} that, for any fixed values of $(m_\theta, f_\theta, \theta_i)$, the final DM abundance is enhanced in our free-kick scenario compared to the conventional one.

It is useful to understand more quantitatively how much the DM abundance can be enhanced.
The DM abundance is determined by the misaligned field value of dilaton and the time at the onset of its oscillation, which in turn depend on the axion kick effect and the dilaton mass, respectively. 
As a result, the DM energy density in our scenario $\rho_\sigma$ versus that in the conventional misalignment scenario $\rho_\theta^{\rm con}$ is roughly 
\bea
\frac{\rho_\sigma}{\rho_\theta^{\rm con}} \sim \left(\frac{m_\theta}{m_\sigma}\right)^{1/2}\left(\frac{f_\theta}{ F}\right)^2 \left(1-\cos\theta_i\right) \, 
\label{eq:DM_relic}
\eea
up to an overall coefficient which will be determined numerically. The details which give rise to the above equation are collected in Appendix~\ref{app:analytic_estimates}.
It suggests that, for any fixed values of $(m_\theta, f_\theta, \theta_i)$, DM relic density is more enhanced when $m_\sigma$ and $F$ are smaller. The reasons are twofold: first, the onset of dilaton oscillation gets delayed with a smaller $m_\sigma$; second, the kick effect is more significant in the same limit.

\begin{figure}[t] 
\centering 
\includegraphics[width=0.45\textwidth]{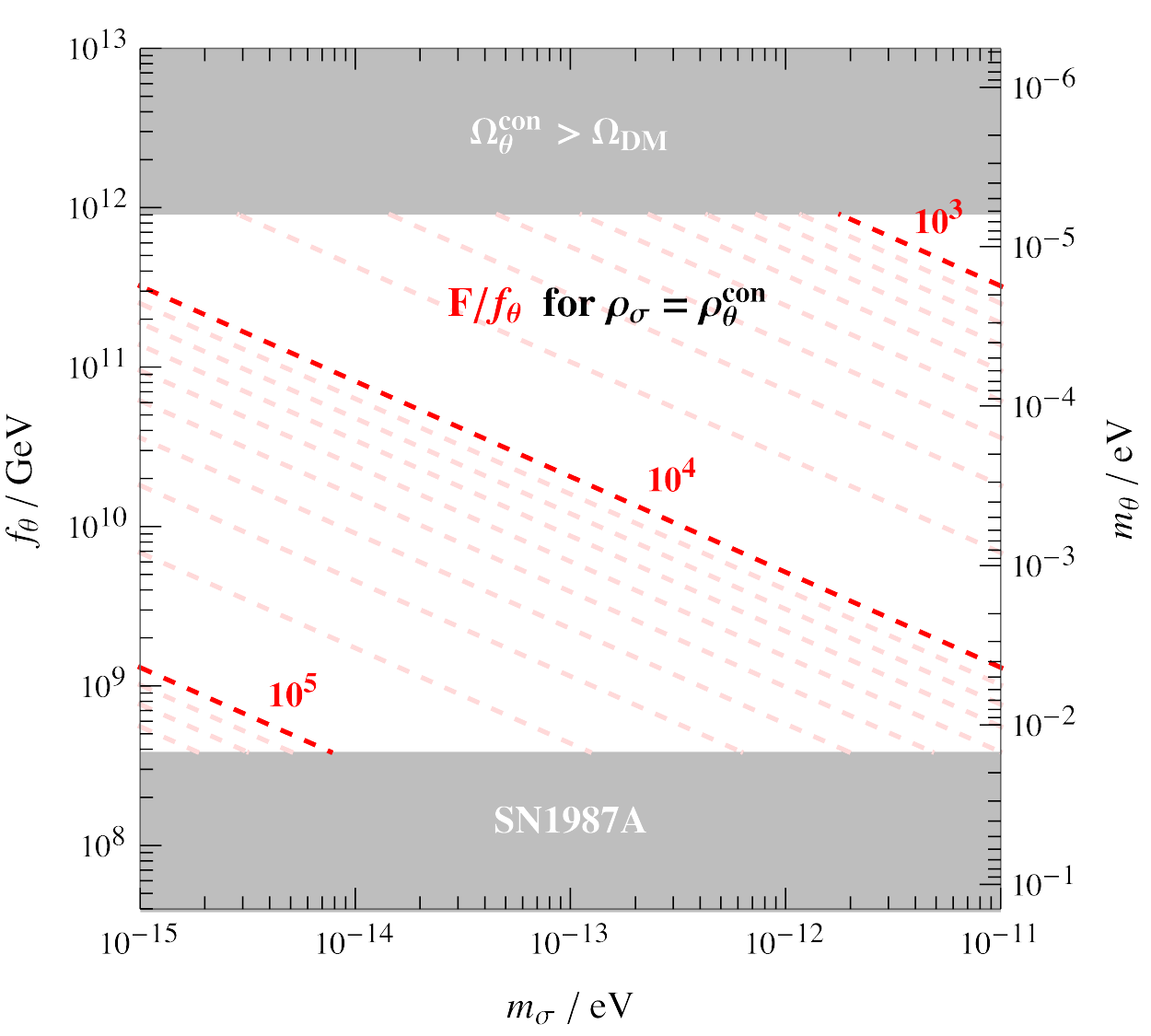} 
\caption{Parameter space of the QCD axion and the light dilaton. The dashed contours indicate the values of $F/f_\theta$ such that $\rho_\sigma = \rho_\theta^{\rm con}$ with $\theta (t_i) \simeq 1$. The upper and lower gray-shaded regions are excluded.}
\label{fig:QCDAxion} 
\end{figure}

We examine the axion free-kick misalignment in the QCD axion scenario, which is strongly motivated as a solution to the strong CP problem~\cite{Peccei:1977hh,Peccei:1977ur,Weinberg:1977ma,Wilczek:1977pj}.
We consider the nontrivial temperature dependence of the QCD axion mass~\cite{Bae:2008ue,Wantz:2009it,Borsanyi:2016ksw,Ballesteros:2016xej,Berkowitz:2015aua,Borsanyi:2015cka,Kitano:2015fla,Petreczky:2016vrs,Taniguchi:2016tjc} above the scale of QCD phase transition $\Lambda_{\rm QCD}\sim 100\,{\rm MeV}$. Besides, the QCD axion mass and decay constant are related roughly as $m_\theta^2 f_\theta^2\sim (80 \text{MeV})^4$. For simplicity, we consider the dilaton mass $m_\sigma < H (\Lambda_{\rm QCD}) \sim 10^{-11}\,{\rm eV}$, such that the dilaton starts to oscillate after QCD phase transition; the other case, i.e. $m_\sigma > H (\Lambda_{\rm QCD}) \sim 10^{-11}\,{\rm eV}$, is left for future work. The parameter space of the QCD axion and the dilaton is shown in Fig.~\ref{fig:QCDAxion}. Assuming $\theta (t_i) \simeq 1$, DM is overproduced for $f_\theta \gtrsim 10^{12}\,{\rm GeV}$ in the conventional misalignment mechanism~\cite{Bae:2008ue,Wantz:2009it,Borsanyi:2016ksw,Ballesteros:2016xej}; see the upper gray-shaded region. When $f_\theta \lesssim 10^{12}\,{\rm GeV}$, the QCD axion does not constitute all the DM, we compute the value of $F/f_\theta$ for each $(m_\sigma,m_\theta)$ such that $\rho_\sigma/\rho_\theta^{\rm con}=1$; see the dashed contours in the unshaded region. Using Eq.~(\ref{eq:DM_relic}), one can estimate how much $\rho_\sigma$ is enhanced compared to $\rho_\theta^{\rm con}$ with lower values of $F$.
The stellar cooling bound from SN1987A~\cite{Raffelt:1996wa,Bionta:1987qt,Kamiokande-II:1987idp,Alekseev:1987ej,Chang:2018rso,Carenza:2019pxu} is illustrated as the lower gray-shaded region.
Although our free-kick mechanism is very different from the axion kinetic misalignment mechanism~\cite{Co:2019jts,Chang:2019tvx}, a similar lowered axion decay constant can be realized.

\vspace{0.1in}
\noindent
\noindent\textbf{Discussion and Outlook.---}
In this Letter, we propose a new mechanism called the axion free-kick misalignment mechanism, where DM abundance is produced due to the initial misalignment of axion, but axion itself is not identified as the main component of DM. Rather, the motion of axion kicks the other ultralight scalar field in the setup, which in this work is dilaton. Eventually, dilaton starts to oscillate, and at the onset of oscillation, the misaligned dilaton field value can be calculated from the kick effect. We find that dilaton dominates over axion in the DM energy density, and total DM abundance today is enhanced compared to the conventional misalignment mechanism, despite the fact that the initial energy densities are the same in our scenario and the conventional scenario. One future direction is to perform the current analysis more systematically and precisely. 

Dilaton phenomenology is also important to confirm or exclude our mechanism. Here we predict a dilaton whose mass is lower than that of an axion but with a larger decay constant. See e.g. Refs.~\cite{Kaplan:2000hh,Damour:2010rp,Arvanitaki:2014faa} for discussions on dilaton phenomenology. Furthermore, similar to pions or axions, precision measurements of dilaton couplings can also shed light on UV physics based on the anomaly matching argument. Anomalous couplings in the effective theory of dilaton can also change the conventional picture phenomenologically~\cite{Csaki:2022htl}. See also~\cite{Sonner:2006yn,Russo:2022pgo} on the interplay between axion and dilaton in cosmology. 

Last but not least, we see the possibility that the cosmological constant (CC) problem can be solved with a dilaton-like particle in the paradigm of cosmological naturalness, where the nonvanishing CC can induce a scale-invariant potential for the dilaton-like particle, which might be used to select the value of CC in the early universe. 

\vspace{0.1in}
\noindent
{\bf Acknowledgements.---}
We thank Marco Peloso for discussions, and FIFA world cup $2022$ which gives us the inspiration for the title.
The work of L.X.X. is supported in part by the MIUR under contract 2017FMJFMW (PRIN2017). 
The work of S.Y. is supported by the research grants: “The Dark Universe: A Synergic Multi-messenger Approach” number 2017X7X85K under the program PRIN 2017 funded by the Ministero dell'Istruzione, Universit\`{a} e della Ricerca (MIUR); “New Theoretical Tools for Axion Cosmology” under the Supporting TAlent in ReSearch@University of Padova (STARS@UNIPD). 
S.Y. is also supported by Istituto Nazionale di Fisica Nucleare (INFN) through the Theoretical Astroparticle Physics (TAsP) project.

\bibliography{axion_dilaton}

\clearpage
\onecolumngrid
\appendix
\makeatletter

\label{supp}

\newpage


\noindent
\section{Analytic Estimation of Dark Matter Abundance in Axion Free-kick Misalignment Mechanism}
\label{app:analytic_estimates}
\setcounter{equation}{0}
\setcounter{figure}{0}
\setcounter{table}{0}
\renewcommand{\theequation}{A\arabic{equation}}
\renewcommand{\thefigure}{A\arabic{figure}}
\renewcommand{\thetable}{A\arabic{table}}

The supplementary material contains the analytic analysis of the equations of motions of axion and dilaton, which are
\bea
&&\ddot{\theta} + \left(3H + 2 \dot{\bar\sigma}\right)\dot{\theta} + m_\theta^2 e^{2\bar\sigma} \sin\theta = 0 \, ,\\
&&\ddot{\bar\sigma} + \left(3H + \dot{\bar\sigma}\right)\dot{\bar\sigma} + \frac{m_\sigma^2}{\epsilon}e^{2\bar\sigma}\left(1-e^{-\epsilon\bar\sigma}\right) + 4\frac{m_\theta^2 f_\theta^2}{F^2}\left(1-\cos\theta\right)e^{2\bar\sigma} - \frac{f_\theta^2}{F^2}\dot{\theta}^2 = 0 \, ,\label{eq:EoMsigma}
\eea
where $\bar\sigma\equiv \frac{\sigma}{F}$ and $\theta$ are the dimensionless dilaton and axion fields normalized by their decay constants $F$ and $f_\theta$, respectively.
The standard Hubble parameter in a radiation-dominated universe is given by 
\bea
H=\frac{\dot{R}(t)}{R(t)}=\frac{\pi \sqrt{g_{\text{eff}}}}{\sqrt{90}} \frac{T^2}{M_{\text{pl}}}\sim \frac{1}{2 t}\ , 
\eea
where the time $t$ and the temperature $T$ are related, $g_{\text{eff}}$ is the effective number of degrees of freedom in the thermal bath, and $M_{\text{pl}}\sim 10^{18}\,{\rm GeV}$ is the Planck scale. One can read off the relation $R(t)\propto t^{1/2}\propto T^{-1}$.

In this work, we assume that the initial velocities of the axion and the dilaton vanish, and we only set sizable initial misalignment value for the axion field, while the dilaton field value is initially near its minimum, i.e. 
\bea
\dot{\theta} (t_i) = \dot{\bar\sigma} (t_i) =0, \ \theta (t_i) \sim \mathcal{O}\left(1\right), \ \bar\sigma (t_i) \sim 0. 
\eea
Therefore, at the initial time $t_i$, the dilaton energy density $\rho_\sigma$ vanishes, only the axion energy density $\rho_\theta$ is nonzero whose value is the same as in the conventional axion misalignment mechanism, for any fixed values of axion mass $m_\theta$ and decay constant $f_\theta$. For simplicity, we will denote $\theta (t_i)$ as $\theta_i$ and the same for other quantities in the following. 
As in the conventional misalignment, dark matter abundance is sourced by the initial misalignment of the axion field. 

Nevertheless, being different from the conventional case, axion itself in our scenario is not dark matter at the end of time evolution. Due to the non-trivial interplay between the axion and the dilaton, the axion starts to oscillate first before the dilaton if
\bea
m_\sigma < m_\theta \ .
\eea
When Hubble parameter drops below $m_\theta$, and the motion of the axion kicks the dilaton through the `kick' term 
\bea
\mathcal{K}(\theta)= \frac{4 m_\theta^2 f_\theta^2 \left(1-\cos[\theta]\right) e^{2 \bar\sigma}- f_\theta^2 \dot{\theta}^2}{F^2}\ .
\eea
Due to the kick effect, the dilaton starts to move away from its minimum and eventually it oscillates when the Hubble parameter drops below $m_\sigma$.
At the end of time evolution, dilaton will dominate the energy density and it is identified as the main component of dark matter. The kick effect is the most significant if $\bar\sigma (t_i) \sim 0$, in this case we call our scenatio the axion `free-kick' misalignment mechanism. Otherwise, the dark matter abundance is also partially sourced by the initial dilaton misalignment, the kick effect is less important. As we will show, the dark matter abundance can be enhanced compared to that in the conventional axion misalignment for fixed values of $m_\theta$ and $f_\theta$.

Other natural parametric choices are $\epsilon \ll 1$ and $\frac{f_\theta^2}{F^2}< 1$, which means that the explicit breaking of scale invariance is a small effect, and the spontaneous-breaking scale of the scale invariance is higher than the Peccei-Quinn (PQ) scale. The kick effect is significant if $\mathcal{K}(\theta) \gg m_\sigma^2$, i.e. $m_\theta^2 f_\theta^2 \left(1-\cos\theta_i\right)\gg m_\sigma^2 F^2$.

There are three main phases in the time evolution, which are characterized by the values of the Hubble parameter and the masses of the axion and the dilaton. 

\noindent
\subsection{$H > m_\theta$}
At first, the Hubble friction is strong enough to trap the axion at its initial field value. Due to the `kick' term $\mathcal{K}(\theta)$, dilaton field is moving toward a negative value. When $\dot{\bar\sigma}\ll H$, the equation of motion of the dilaton is approximately 
\bea
\ddot{\bar\sigma} + 3H\dot{\bar\sigma}  + 4\frac{m_\theta^2 f_\theta^2}{F^2}\left(1-\cos\theta_i\right) = 0 \ ,
\eea
which can be rewritten as
\bea
\ddot{A} +\frac{3}{16t^2} A + \left(\frac{R}{R_0}\right)^{3/2} 4\frac{m_\theta^2 f_\theta^2}{F^2}\left(1-\cos\theta_i\right)=0 \, 
\eea
using the redefined dilaton field $A$ as
\bea
A = \left(\frac{R}{R_0}\right)^{3/2} \bar\sigma\ .
\eea
By straightforward calculation, we find the dilaton field value as
\bea
\left. \bar\sigma \right|_{H > m_\theta} \simeq -t^2\frac{4}{5}\frac{m_\theta^2 f_\theta^2}{F^2}\left(1-\cos\theta_i\right)\, \ ,
\eea
and accordingly the dilaton velocity as
\bea
\left| \dot{\bar\sigma} \right| \sim \frac{1}{t} \left(m_\theta^2 t^2 \frac{f_\theta^2}{F^2}\right) \sim H \left(m_\theta^2 t^2 \frac{f_\theta^2}{F^2}\right)\ .
\eea
Since $m_\theta t \ll 1$ and $f_\theta^2 \ll F^2$, we find $\left| \dot{\bar\sigma} \right|\ll H$. This justifies the assumption that $\dot{\bar\sigma}\ll H$, which we impose at the beginning of the calculation.

\noindent
\subsection{$m_\sigma < H < m_\theta$}

When $m_\sigma < H < m_\theta$, the axion start to oscillate with the frequency characterized by its mass $\sim m_\sigma^{-1}$, which is much faster than the typical Hubble time $H^{-1}$. 
Therefore, when considering the effect of the axion oscillation on the dilaton evolution, one can take the average 
\bea
\left<\mathcal{K}(\theta)\right>=\left< 4\frac{m_\theta^2 f_\theta^2}{F^2}\left(1-\cos\theta\right) - \frac{f_\theta^2}{F^2}\dot{\theta}^2\right> \simeq 2\frac{m_\theta^2 f_\theta^2}{F^2}\left(1-\cos\theta_i\right) \left(\frac{t_{\theta}}{t}\right)^{3/2} \, ,
\eea
where the last factor accounts for the redshift for matter-like oscillation, whose onset time $t_{\theta}$ is given by
\bea
\frac{3}{2} H_{\theta} = \frac{3}{4t_{\theta}} \sim m_\theta \, .
\eea
The equation of motion of the dilaton is approximately given by
\bea
\ddot{A} +\frac{3}{16t^2} A + \left(\frac{t}{t_{\theta}}\right)^{-3/4} 2\frac{m_\theta^2 f_\theta^2}{F^2}\left(1-\cos\theta_i\right) = 0  \, ,
\eea
where $R(t)\propto t^{1/2}$ is considered. 
By calculation, we find
\bea
\left. \bar\sigma \right|_{m_\sigma < H < m_\theta} \simeq -t^{1/2}t_{\theta}^{3/2} 4\frac{m_\theta^2 f_\theta^2}{F^2}\left(1-\cos\theta_i\right)+\frac{16}{5} t_{\theta}^{2} \frac{m_\theta^2 f_\theta^2}{F^2}\left(1-\cos\theta_i\right)  \, ,
\eea
where the constant term is determined by matching the field value of $\bar\sigma$ at the time $t_\theta$. 

\noindent
\subsection{$H< m_\sigma $}
When the Hubble parameter drops below the dilaton mass, i.e. at
\bea
\frac{3}{2} H_{\sigma} = \frac{3}{4t_{\sigma}} \sim m_\sigma \, ,
\eea
the dilaton starts to oscillate with the field value roughly being
\bea
\left. \bar\sigma \right|_{t_{\sigma}} &\simeq& -t_\sigma^{1/2}t_\theta^{3/2}4\frac{m_\theta^2 f_\theta^2}{F^2}\left(1-\cos\theta_i\right)+\frac{16}{5} t_{\theta}^{2} \frac{m_\theta^2 f_\theta^2}{F^2}\left(1-\cos\theta_i\right)\nonumber\\
 &\simeq& - \frac{9}{4}\left(\frac{m_\theta}{m_\sigma}\right)^{1/2}\left(\frac{f_\theta}{F}\right)^2 \left(1-\cos\theta_i\right) +\frac{9}{5}\left(\frac{f_\theta}{F}\right)^2 \left(1-\cos\theta_i\right)\, ,
\label{eq:initialDilaton}
\eea
where the first term dominates due to the ratio of $m_\theta/m_\sigma$.
Eventually, the energy density of the dilaton is determined by 
\bea
\rho_\sigma \simeq 2 m_\sigma^2 F^2 \left(\left. \bar\sigma \right|_{t_{\sigma}}\right)^2 \left(\frac{t_\sigma}{t}\right)^{3/2} \, ,
\eea
where the last factor accounts for the redshift for matter-like oscillation. Notice that this estimate is accurate if the misalignment of the dilaton is not too large at the onset of its oscillation, i.e. $\left. \bar\sigma \right|_{t_{\sigma}}<\mathcal{O}(1)$, or more specifically,
\bea
\frac{9}{4}\left(\frac{m_\theta}{m_\sigma}\right)^{1/2}\left(\frac{f_\theta}{F}\right)^2 \left(1-\cos\theta_i\right) \ll 1 \, .
\eea
This condition also ensures that $\dot{\bar\sigma} \ll H$ during the second stage of the dilaton evolution, and the self-consistency of Eq.~(\ref{eq:initialDilaton}) is justified.

Compared to the conventional axion misalignment mechanism, where the axion itself is dark matter and the DM energy density is
\bea
\rho_\theta^{\rm con} = 2 m_\theta^2 f_\theta^2 \left(1-\cos\theta_i\right)  \left(\frac{t_\theta}{t}\right)^{3/2} \,,
\eea
we find
\bea
\frac{\rho_\sigma}{\rho_\theta^{\rm con}} \sim \left(\frac{m_\theta}{m_\sigma}\right)^{1/2}\left(\frac{f_\theta}{ F}\right)^2 \left(1-\cos\theta_i\right) \, 
\label{eq:DM_relic_supp}
\eea
up to an order one coefficient. 
For fixed values of the axion mass $m_\theta$, the decay constant $f_\theta$ and the initial misalignment $\theta_i$, the dark matter abundance can be enhanced in the `free-kick' scenario depending on the dilaton mass $m_\sigma$ and the decay constant $F$. In other words, due to the interplay between the axion and the dilaton, the dark matter abundance can be reproduced with a lower axion decay constant $f_\theta$, which is interesting for all the axion detection experiments.

\begin{figure}[t] 
\centering 
\includegraphics[width=0.45\textwidth]{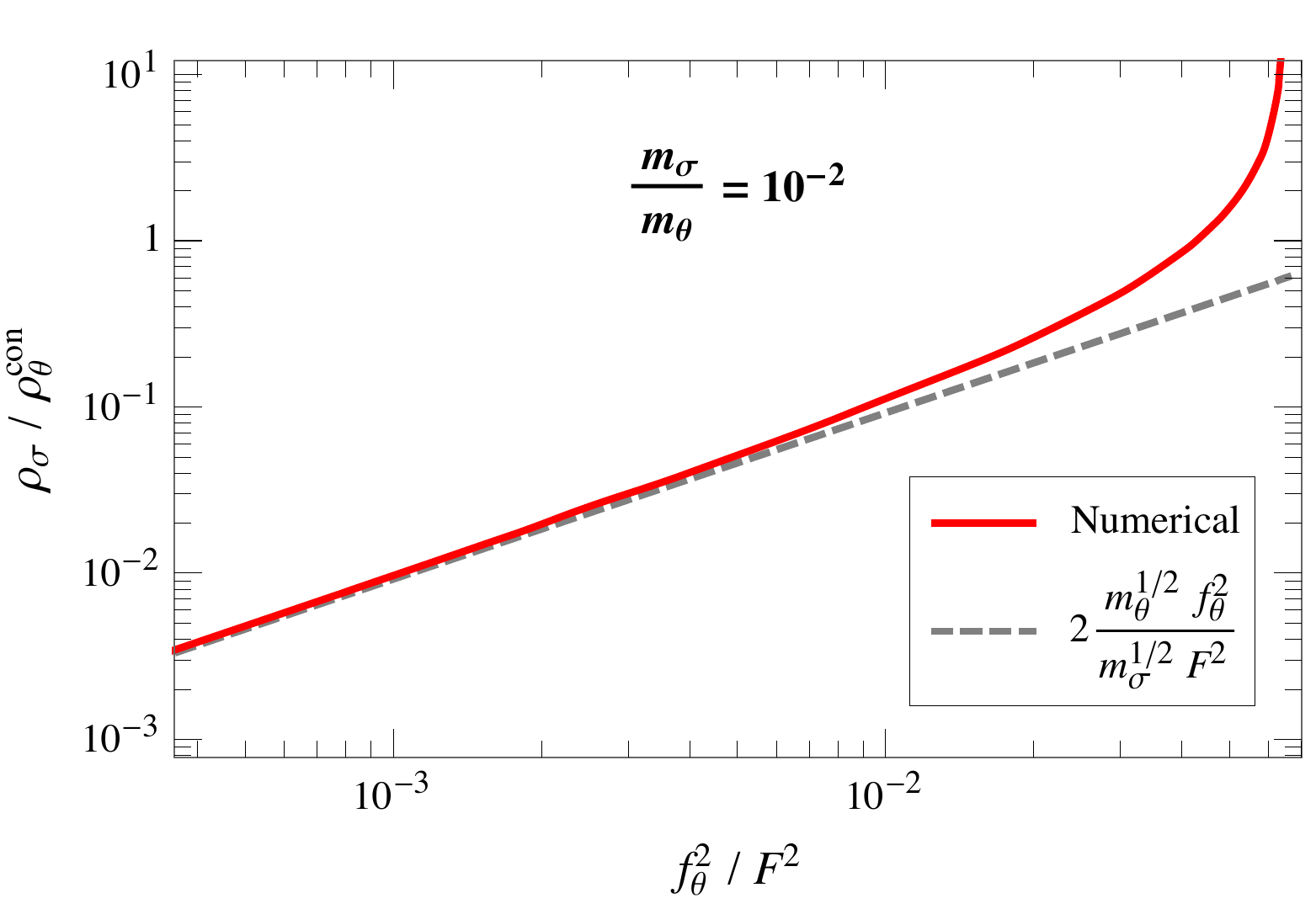} 
\caption{Enhancement of DM abundance in the axion free-kick misalignment mechanism compared to that in the conventional scenario. Here we compare the analytic expression Eq.~(\ref{eq:DM_relic_supp}) with numerical calculation with $m_\sigma/m_\theta$ being fixed. }
\label{fig:ratio} 
\end{figure}

The above analytic estimation gives the semi-quantitative result, and we verify it with more detailed numerical analysis by directly solving Eqs.~(\ref{eq:eom_axion}) and~(\ref{eq:eom_dilaton}); see in Fig.~\ref{fig:ratio}. One can see that Eq.~(\ref{eq:DM_relic}) is only quantitatively accurate when $\rho_\sigma/\rho_\theta^{\rm con}$ is small, but it is still qualitatively correct even when $\rho_\sigma/\rho_\theta^{\rm con}$ becomes large. 
One can repeat the above calculation for QCD axion after taking into account the non-trivial temperature dependence of the QCD axion mass.

\end{document}